\documentclass[twocolumn,preprintnumbers,prl]{revtex4}
\usepackage{physics}
\usepackage{graphicx}
\usepackage{dcolumn}
\usepackage{bm}
\usepackage{color}
\usepackage{siunitx}
\usepackage{float}

\begin{document}

\title{Narrowband Frequency-Entangled Photon Source for Hong-Ou-Mandel Interferometry}
%\title{Narrowband Frequency-Entangled Photon Source for Long-Dynamic-Range Hong-Ou-Mandel Inteferometric Sensing}

\author{Yen-Ju Chen$^{1,2}$}
\email{yen-ju.chen@fau.de}

\author{Sheng-Hsuan Huang$^{1,2}$}
\author{Thomas Dirmeier$^{1,2}$}
\author{Kaisa Laiho$^3$}
\author{Dmitry V. Strekalov$^2$}
\author{Andrea Aiello$^2$}
\author{Gerd Leuchs$^{1,2}$}
\author{Christoph Marquardt$^{1,2}$}

\affiliation{$^1$Department of Physics, Friedrich-Alexander-Universität Erlangen-Nürnberg, Staudtstrasse 7/B2, 91058 Erlangen, Germany.}
\affiliation{$^2$Max Planck Institute for the Science of Light, Staudtstrasse 2, 91058 Erlangen, Germany.}
\affiliation{$^3$German Aerospace Center (DLR e.V.), Institute of Quantum Technologies, Wihelm-Runge-Str. 10, 89081 Ulm, Germany.}

\begin{abstract}
Hong-Ou-Mandel (HOM) interferometry with entangled photons exhibits distinctive quantum features. 
%and has been shown to provide greater robustness against background noise and loss compared to classical optical interferometry. 
By introducing frequency entanglement (discrete-color entangled states) into HOM interference, the characteristic HOM dip is modulated by sinusoidal fringes, which significantly enhance the sensitivity of HOM sensors.
%The introduction of frequency entanglement (discrete color entangled states) modulates the HOM dip with sinusoidal fringes, producing quantum beating that enhances the sensitivity of HOM-based sensors.
%When frequency entanglement (discrete color entangled state) is introduced, the HOM dip becomes modulated by sinusoidal fringes. This quantum beating enhances the sensing capabilities of HOM-based sensors, offering new opportunities for precision measurement applications.
%The full potential of quantum-enhanced sensing can be further explored by integrating quantum interferometric sensors with quantum memories, where narrowband photon pairs are essential for efficient light–matter coupling.
The frequency-entangled photon sources demonstrated to date rely on non-resonant parametric down-conversion (PDC), which limits the photon coherence length and, consequently, restricts the sensing dynamic range to the sub-millimeter scale. 
%Hong-Ou-Mandel (HOM) interferometry with entangled photons exhibits distinctive quantum features. Introducing frequency entanglement (discrete color entangled states) modulates the HOM dip with sinusoidal fringes, producing quantum beating that enhances the sensitivity of HOM-based sensors and opens new avenues for precision measurement. The full potential of quantum-enhanced sensing can be further unlocked by integrating HOM interferometric sensors with quantum memories, where narrowband photon pairs are crucial for efficient light–matter coupling.
In this work, we demonstrate narrowband frequency-entangled photon source based on resonant PDC in a crystalline whispering gallery mode resonator. 
The MHz-level spectral bandwidth of photons enables a meter-scale dynamic range.
With highly nondegenerate frequency-entangled photon pairs featuring a 96 THz frequency detuning, we observe high-contrast quantum beating with sub-picosecond resolution in the HOM experiment.
%Using highly nondegenerate entangled photon pairs with a 96 THz frequency detuning and MHz-level spectral bandwidth, we observe high-contrast quantum beating in the HOM experiment across a wide temporal window, enabled by the meter-scale coherence length of the photon pairs.
%High-contrast quantum beating over a broad temporal window, enabled by the long coherence time of photon pairs, is observed in the HOM experiment using highly-nondegenerate narrowband entangled photon pairs with a 96 THz frequency separation and MHz-level spectral bandwidth.
Our WGMR-based frequency-entangled photon source has potential applications in quantum metrology and quantum information processing.

\end{abstract}

\maketitle

%\textit{Introduction} 

Hong-Ou-Mandel (HOM) interferometry has emerged as a useful tool for the development of novel quantum sensing techniques, with applications ranging from quantum optical coherence tomography \cite{Mago12}, quantum microscopy \cite{Ndagano22}, entangled-photon spectroscopy \cite{Eshun21}, to quantum gyroscopy \cite{Restuccia19}. 
%When two indistinguishable photons enter a balanced beam splitter, the destructive interference of their probability amplitudes suppresses the coincidence detection events at different output ports of the beam splitter. This HOM effect is used in conventional HOM sensors to probe optical time delays with resolution down to the femtosecond scale. Furthermore, by exploiting maximum likelihood estimation strategy, time-delay measurements with attosecond-level resolution have been demonstrated \cite{Lyons18}. 
Interestingly, using frequency-entangled photon pairs in HOM interference results in a characteristic HOM dip modulated by sinusoidal fringes, whose beating period corresponds to the frequency detuning between the signal and idler photons. The optical time delay information can then be extracted from the fringe phase rather than the HOM dip position, thus significantly increasing the measurement resolution.
Leveraging this enhanced information per photon, the quantum beat note reduces the acquisition time required to achieve high-resolution sensing and enabling the precision to approach the quantum Cramér–Rao (QCR) bound. 
Utilizing frequency-entangled photon pairs, a fast HOM sensor operating with a second-scale acquisition time was reported in \cite{Lualdi25}, demonstrating the potential of entanglement-based HOM sensor for real-time quantum sensing. 

The power of quantum sensing may be further unlocked by combining the entanglement-based interferometry with quantum memories.
Typically, the measurement resolution depends on the phase accumulation time between the sensing photon and the target system. Long-lived quantum memories thus offering a promising way to extend the interaction time and enable hybrid sensing applications \cite{Zaiser16,Ahmadi25}. 
%For instance, an optical interferometric velocimeter with enhanced sensitivity based on atomic quantum memory was demonstrated in \cite{Ahmadi25}. 
%The full potential of quantum-enhanced sensing may be further unlocked by combining the entanglement-based interferometry with quantum memories, where the narrowband entangled photon sources are crucial for achieving efficient interfacing.
To achieve efficient interfacing between photons and quantum memories, narrowband entangled photon sources are essential for fully harnessing the potential of quantum-enhanced sensing. 
%To fully harness the potential of quantum-enhanced sensing, narrowband entangled photon sources are essential for efficiently interfacing entangled photons with quantum memories.

To date, the frequency-entangled photon source has been realized in Sagnac loop \cite{Ramelow09, Torre23} and two-crystal configuration \cite{Lualdi25,Chen19}. 
Although previous approaches have achieved frequency detuning up to 144 THz, the relatively broad photon bandwidths ($\approx$ 0.5 nm) correspond to coherence times of only a few picoseconds, limiting the dynamic range of HOM sensor to the sub-millimeter scale ($\approx$ 0.76 mm).
Moreover, the broad spectral bandwidth of photons generated in bulk nonlinear media poses challenges for efficient light–matter interactions, which typically require MHz-level bandwidths \cite{Zhang12,Wu17}. 
%However, the relatively short coherence time of photon pairs from bulk nonlinear media limits the HOM sensing dynamic range to a few millimeters, while their broad spectral bandwidth poses challenges for achieving efficient light-matter interactions, which typically require MHz-level photon bandwidths \cite{Zhang12,Wu17}.

To overcome the coherence time and spectral bandwidth limitations, resonant parametric down-conversion (PDC) provides a powerful approach to generate narrowband photon pairs with significantly extended temporal coherence. 
While achieving multi-resonance conditions under large frequency detuning in resonant PDC is generally challenging, whispering gallery mode resonators (WGMRs) offer a particularly advantageous platform \cite{Otterpohl19,Förtsch13,Huang24}. Taking advantage of the total internal reflection of light within the WGMR, the triple resonance of pump, signal and idler fields at significantly different wavelengths can be naturally supported without the need for complex reflector fabrication or active cavity control. 
%In particular, PDC in whispering gallery mode resonators (WGMRs) offers an efficient and versatile platform for nonclassical light generation, including squeezed light \cite{Otterpohl19} and heralded single photons \cite{Förtsch13}.
%has been demonstrated to be an efficient and versatile source for generating nonclassical light, such as squeezed light \cite{Otterpohl19} and heralded single photons \cite{Förtsch13}. 
%Benefit from the high quality factor (Q) and strong optical field confinement within WGRMs, these resonators can generate bright photon pairs with subnatrual linewidths and can operate at extremely low pump power \cite{Huang24}. 

In this work, we demonstrate a narrowband frequency-entangled photon source based on PDC in a crystalline WGMR. Using highly-nondegenerate frequency-entangled photon pairs with a frequency detuning of 96 THz, we observed high-contrast HOM interference fringes with sub-picosecond resolution over the wide photon coherence length.
The MHz-level spectral bandwidth enables a significantly extended dynamic range reaching the meter scale.
This long dynamic range relaxes the requirement for precise alignment of the interferometer paths, which may have advantages for long-baseline measurements \cite{Mohageg22}.
%Taking advantage of the total internal reflection of light within WGMR, the triple resonance of pump, signal and idler fields at significantly different wavelengths can be achieved without complex reflector fabrication and cavity control. 
%Moreover, the bandwidth of photon pairs can be efficiently tuned by controlling the coupling condition between the WGMR and the prism coupler. This ability to flexibly control the bandwidth of photon pairs highlights the versatility of our WGMR-based photon source for applications requiring adjustable spectral properties \cite{Zhang12}\cite{Huang25}. 
Our WGMR-based entangled photon source may have potential applications in the study of entangled light-matter interactions and quantum metrology.

%\textit{Theory} 

\begin{figure}[!t]
\centering
\includegraphics[width=8.5cm]{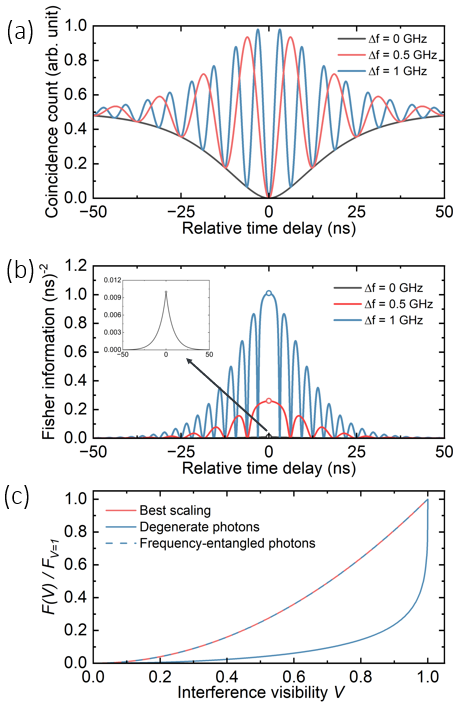}
\caption{\label{fig:theory} (a) HOM interference of frequency-entangled photons and the corresponding (b) Fisher information for various frequency detuning between the signal and idler photons $\Delta f=\Delta\omega/2\pi$. The open circle represents the point where the Fisher information is undefined. The inset shows an enlarge view of the Fisher information calculated with degenerate photon pairs. The interference visibility in the calculations is $V=1$. (c) Fisher information ratio as a function of interference visibility. The angular frequency detuning between the signal and idler photons for frequency-entangled state is $\Delta\omega=10$ THz. In the calculations, the waveform of the state is double-exponential decay function with cavity decay rates $\Gamma_{s}=\Gamma_{i}=$ 1/20 ns$^{-1}$.
}
\end{figure}

The frequency-entangled state can be described as

\begin{equation}
\begin{aligned}
    \label{eq:freqEnt}
    \ket{\psi}&=\frac{1}{\sqrt{2}}\iint dtdt' e^{-i(\omega_{s}t+\omega_{i}t')}\phi(t-t') \\
    &\times[a^{\dagger}_{sA}(t)a^{\dagger}_{iB}(t')+a^{\dagger}_{sB}(t)a^{\dagger}_{iA}(t')]\ket{0} ,
\end{aligned}  
\end{equation}

\noindent where $\phi(t-t')$ is the wave-packet temporal profile of photon pair.
The bosonic operator creating a signal (idler) photon at time $t$ in the input port A (B) of the HOM interferometer is denoted as $a^{\dagger}_{sA (iB)}(t)$. 
%$a^{\dagger}_{sA (iB)}$ is the bosonic operator creating a signal (idler) photon in the input port A (B) of the HOM interferometer. 
The angular frequencies of the signal and idler photons are denoted as $\omega_{s}$ and $\omega_{i}$, respectively.
After the 50/50 beam-splitter transformation, the Glauber correlation function between the two output ports, mode C and D, is calculated as 

\begin{equation}
\begin{aligned}
    \label{eq:G2def}
    G^{(2)}(\tau)=\langle a^{\dagger}_C(t_0+\tau)a^{\dagger}_D(t_0)a_D(t_0)a_C(t_0+\tau) \rangle,
\end{aligned}  
\end{equation}

By integrating the Glauber correlation function over the coincidence window, the coincidence detection probability as a function of the time difference $\tau$ between the photons arriving at the two output ports can be expressed as \cite{Lualdi25}

\begin{equation}
    \label{eq:Pc}
    P_{c}=\frac{1}{2}[1-V\int d\tau \phi(\tau+\Delta t)\phi(\tau-\Delta t)\cos{(\Delta\omega\Delta t)}],
\end{equation}

\noindent where $V$ is the interference visibility. 
The relative time delay between the two arms of the interferometer is denoted as $\Delta t$. The angular frequency detuning between the signal and idler is defined as $\Delta\omega=\omega_{s}-\omega_{i}$. See supplementary material for details. 
In the resonant PDC case, the wave packet of the photon pair in the time domain can be described by a double-exponential decay function \cite{Chuu11}

\begin{equation}
    \label{eq:wavePacket}
    \phi(t_{s}-t_{i})=\sqrt{\frac{2\Gamma_s\Gamma_i}{\Gamma_s+\Gamma_i}} \cross
    \begin{cases}
    e^{-\Gamma_s (t_{s}-t_{i})},&  t_{s}-t_{i} \geq 0, \\
    e^{\Gamma_i (t_{s}-t_{i})},&  t_{s}-t_{i} < 0. 
    \end{cases}
\end{equation}

The calculated HOM interference patterns for various frequency detuning between the signal and idler photons are shown in Fig. \ref{fig:theory}(a). In the degenerate case, the interference exhibits a characteristic HOM dip, whose width scales with the coherence time of the photon pairs. This broad HOM dip limits the temporal resolution for time-resolved detection applications. 
As the frequency detuning between the signal and idler photons increases, the resulting HOM dip is modulated by sinusoidal fringes, with a fringe period inversely proportional to the frequency detuning between the two photons. This modulation transforms the broad HOM dip into a finely structured interference pattern, which is highly sensitive to changes in relative time delay. The temporal resolution of the time-delay measurement is thus significantly enhanced.
The time-delay information extracted from the HOM interference can be evaluated by analyzing the Fisher information (FI).  
Considering negligible photon loss, the FI at relative time delay $\Delta t$ can be written as \cite{ Lyons18,Chen19}

\begin{figure*}[!tb]
\centering
\includegraphics[width=17.5cm]{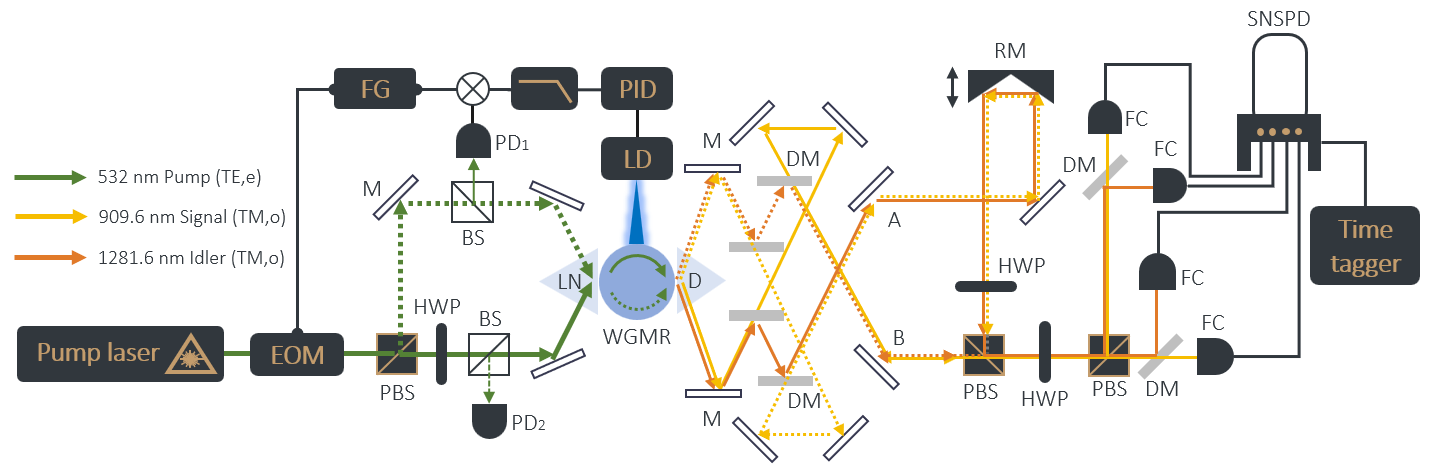}
\caption{\label{fig:setup} Schematic of the experimental setup. EOM: electro-optic modulator, FG: function generator, M: mirror, PBS: polarizing beam splitter, HWP: half-wave plate, BS: non-polarizing beam splitter, PD: photodetector, PID: proportional–integral–derivative controller, LD: laser diode, LN: x-cut LiNbO$_{3}$ prism, WGMR: whispering gallery mode resonator, D: diamond prism, DM: dichroic mirror, RM: retroreflector mirror, FC: fiber collimator, SNSPD: superconducting nanowire single-photon detector, TM: transverse magnetic modes, TE: transverse electric modes, o: ordinary, and e: extraordinary.
}
\end{figure*}

\begin{equation}
    \label{eq:Fisher}
    F=\frac{[\partial_{\Delta t}P_{c}(\Delta t)]^{2}}{P_c(\Delta t)}+\frac{[\partial_{\Delta t}P_{a}(\Delta t)]^{2}}{P_a(\Delta t)}=\frac{[\partial_{\Delta t}P_{c}(\Delta t)]^{2}}{P_c(\Delta t)[1-P_c(\Delta t)]},
\end{equation}

\noindent where $P_{a}=1-P_{c}$ denotes the detection probability of anticoincidence events. 
The calculated FI is shown in Fig.~\ref{fig:theory}(b). As the FI scales quadratically with the frequency detuning between the signal and idler photons, the measurement precision can be substantially improved when using highly nondegenerate frequency-entangled photon pairs. Additionally, the use of narrowband photon pairs significantly extends the available sensing range with high Fisher information.

In the ideal case, the FI from the perfect HOM interference saturates the QCR limit. In practice, however, the precision of a HOM sensor is limited by imperfect interference visibility, which significantly reduces the attainable FI. In this scenario, the waveform of the state, which is related to its effective occupation area in the time-frequency phase space, offers an alternative way to improve the measurement precision. 
As shown in Fig. \ref{fig:theory}(c), for degenerate photon pairs with a double-exponential decay waveform, the ratio of the maximal achievable FI at non-ideal visibility to its ideal value falls to 0.5 when the visibility is reduced to 0.98.
Remarkably, the frequency-entangled state shows robust precision improvements against the reduced visibility, and saturates the best scaling of the FI ratio, which scales quadratically with the visibility $F(V)=V^{2}F_{V=1}$ \cite{Meskine24}. Without the need for additional spectral mode engineering, the frequency-entangled state exhibits sinusoidal characteristics in its chronocyclic Wigner function similar to those of a Schrödinger-cat-like state, rendering the frequency-entangled state inherently more resilient to visibility imperfections.

%\textit{Experimental setup and results} 

\begin{figure}[!t]
\centering
\includegraphics[width=8.5cm]{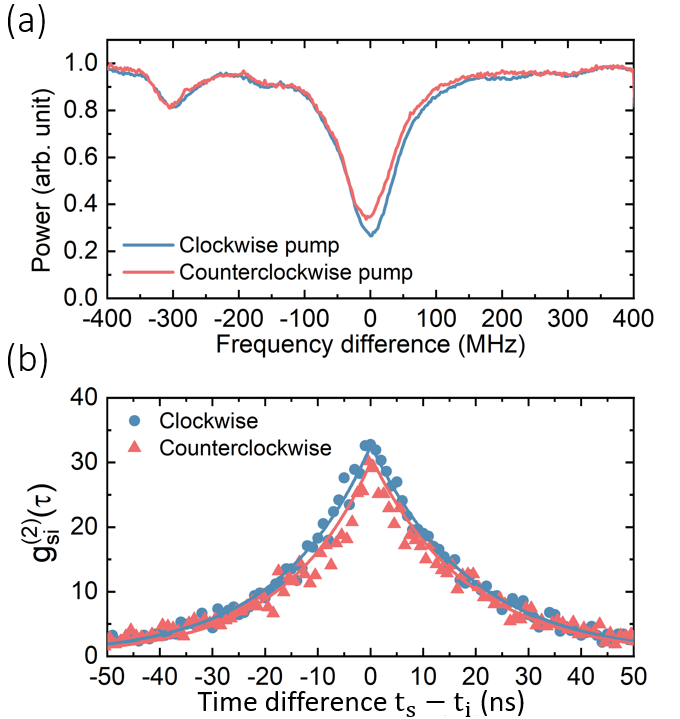}
\caption{\label{fig:source}(a) Reflectance spectra for the pump beams in the CW and CCW directions measured from photodetectors $D_{1}$ and $D_{2}$, respectively. (b) Glauber correlation functions of photon pairs generated in the CW and CCW directions of the WGMR.  
}
\end{figure}

The schematic of the experimental setup is shown in Fig. \ref{fig:setup}. Our WGMR is made of z-cut 5\% MgO-doped lithium niobate (LiNbO$_{3}$) with a resonator radius $R\approx0.745$ mm and a rim radius $r\approx0.144$ mm. The temperature of the WGMR is stabilized at 22.5 $^{\circ}$C to satisfy type-I phase-matching condition, with the pump, signal and idler wavelengths at 532 nm, 909.6 nm and 1281.6 nm, respectively. 
By shining a 462 nm heating laser onto the WGMR, the resonator temperature is further fine-tuned to lock the pump mode resonant frequency to the pump laser frequency based on the Pound-Drever-Hall (PDH) technique. 
The WGMR is pumped with a continuous-wave frequency doubled Nd:YAG laser from the clockwise (CW) and counterclockwise (CCW) directions via a x-cut LiNbO$_{3}$ prism. 
Non-polarizing beam splitters are used to extract the reflected beams, which are then measured with photodiodes to obtain the reflectance spectra.
To ensure that the pump beams from two directions are coupled to the same resonant mode, both beams are aligned to produce the same reflected spectra, and the reflectance is optimized to achieve higher down-conversion efficiency. 
The reflectance spectra for the pump in the CW and CCW directions are shown in Fig. \ref{fig:source}(a). The measured bandwidth is 99.1 MHz, corresponding to a quality factor (Q) of $5.7\times10^{6}$. 

The generated signal and idler photons are coupled out by a diamond prism. These highly nondegenerate photon pairs are first separated by a dichroic mirror (DM). Then, the signal (idler) photons from the CW (CCW) mode are overlapped with the idler (signal) photons from the CCW (CW) mode on a dichroic mirror to prepare the frequency-entangled state. 
After the entangled state preparation, the beams are recombined into the same path with orthogonal polarization modes by using a half-wave plate (HWP) and a polarizing beam splitter (PBS). The HOM interference can then be efficiently implemented by applying a diagonal polarization transformation with a HWP, followed by beam separation using a PBS. 
After the interference, a DM with a 1000 nm cut-off wavelength is used to separate the signal and idler photons prior to detections. Under destructive interference, the coincidence count can only be measured between the detectors located at the same output port of the PBS. 
Note that the pump laser coupled out of the WGMR is filtered out by using bandpass filters before the fiber collimators (not shown in the figure) to suppress accidental coincidence counts.

The measured Glauber correlation functions of the photon pairs generated in CW and CCW modes are shown in Fig. \ref{fig:source}(b). The cavity decay rates of the signal and idler photons are $\Gamma_{s}=1/18.84$ ns$^{-1}$ and $\Gamma_{i}=1/17.53$ ns$^{-1}$, corresponding to single-photon bandwidths of 8.45 MHz and 9.08 MHz, respectively.
The bandwidth of photon pairs can be calculated as $\Delta\Omega = [(\sqrt{\Gamma_{s}^{4}+6\Gamma_{s}^{2}\Gamma_{i}^{2}+\Gamma_{i}^{4}}-\Gamma_{s}^{2}-\Gamma_{i}^{2})/2]^{1/2}$, yielding a photon-pair bandwidth of 5.63 MHz \cite{Chuu11}. This high-Q WGMR enables high source brightness, with a detection rate of the frequency-entangled photons of $1.1\times10^{6}$ pairs per second per mW pump power. 

\begin{figure}[!t]
\centering
\includegraphics[width=8.5cm]{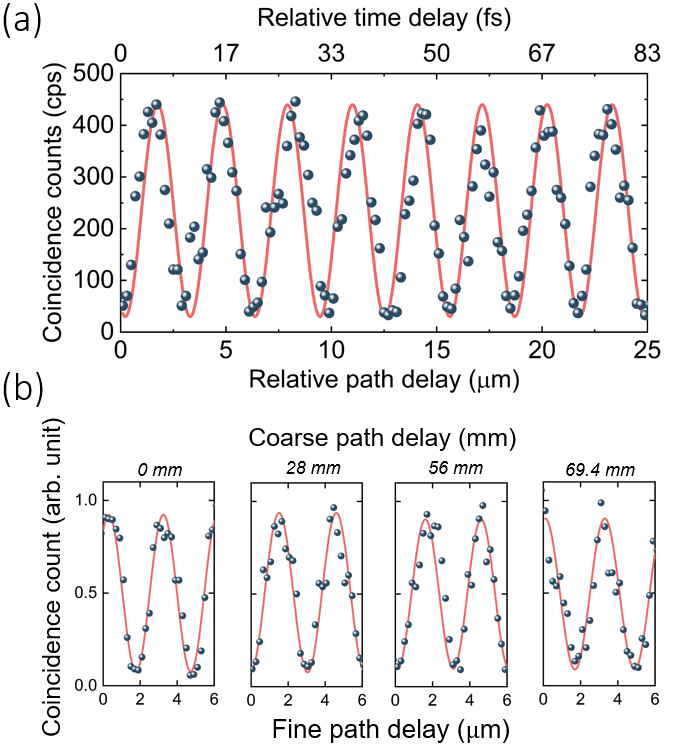}
\caption{\label{fig:qbeating} (a) HOM interference of narrowband frequency-entangled photons. 
The highest interference visibility achieved is 87.3$\%$ based on the best fit (red curve). 
(b) HOM interference with various optical path delays. The HOM interference fringes are measured under critical coupling conditions, where the coupling rate is approximately equal to the internal loss rate. Under these conditions, the photon-pair bandwidth is approximately 5.6 MHz.
The red curves are the best fits to the measured data.}
\end{figure}

To probe the HOM interference properties of the frequency-entangled photon pairs, a retroreflector mirror (RM) mounted on a high-precision linear translation stage is employed to introduce a controllable optical path delay between the two arms of the HOM interferometer. 
By summing the coincidence counts between two output ports of the interferometer within a coincidence window ($\pm50$ ns), the measured coincidence count exhibits sinusoidal oscillations [Fig. \ref{fig:qbeating}(a)]. The fitted beating period is 3.08 $\mu$m, which is close to the expected period of 3.13 $\mu$m calculated from the 95.66 THz frequency detuning between the signal (909.6 nm) and idler photons (1281.6 nm). The interference visibility is $87.3 \%$, and can be further improved by optimizing the CW and CCW pump power \cite{Huang25}. 
Due to the long coherence length of the photon pairs, our HOM interferometer does not require precise initial calibration of the relative optical path delay to locate the HOM dip. Moreover, the broad temporal region with nonzero interference visibility provides a wide dynamic range for sensing applications, where the absolute distance measurement can be performed by counting the fringes in a manner of vernier. 
To demonstrate this capability, we measure the HOM interference at various relative path delays [Fig. \ref{fig:qbeating}(b)]. At the limit of our translation stage, the interference still exhibits fringes with high visibility. 
Note that due to mechanical vibrations in the experimental setup, a slight phase drift in the interference fringes can be observed during the measurements. The maximum relative path delay demonstrated here remains small compared to the meter-scale coherence length of narrowband photon pairs generated by the WGMR, the interference visibility therefore shows no significant difference at different path delays. Nevertheless, the demonstrated optical delay already exceeds the coherence length achievable with non-resonant PDC sources, which is typically limited to a few millimeters \cite{Lualdi25}. 

\begin{figure}[t]
\centering
\includegraphics[width=8.5cm]{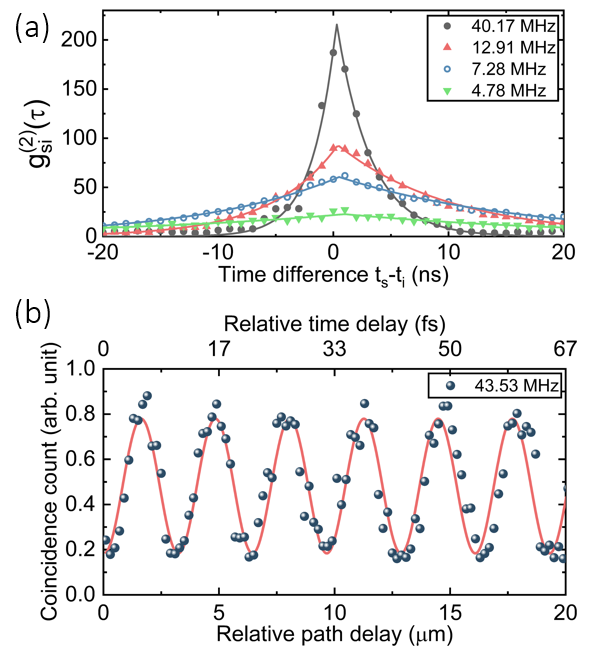}
\caption{\label{fig:over} (a) Glauber correlation functions with various bandwidths of photon pairs $\Delta\Omega$. (b) HOM interference measured under the overcoupling condition exhibits sinusoidal oscillations with a fitted beating period of 3.20 $\mu$m and an interference visibility of 72.6$\%$. The photon-pair bandwidth is approximately 43.53 MHz.}
\end{figure}

In our experimental configuration, the generated photons are evanescently coupled out of the WGMR using a diamond prism coupler. By precisely adjusting the distance between the prism and the WGMR, the coupling rate can be broadly and continuously tuned, providing an efficiently and flexible way of controlling the bandwidth of the photon pairs. 
When the prism is moved close to the WGMR, the coupling rate increases to exceed the internal loss rate. This overcoupling condition leads to a higher total cavity decay rate, resulting in a broader bandwidth of the emitted photon pairs. Conversely, when the prism is positioned farther from the WGMR, the coupling rate decreases, allowing photons to propagate longer within the resonator. The photon lifetime is therefore extended, leading to a lower cavity decay rate and narrower bandwidth of photon pairs. 
The measured Glauber correlation functions for difference coupling distances of the WGMR are shown in Fig. \ref{fig:over}(a). The photon-pair bandwidth demonstrated can be continuously tuned from 4.78 MHz to 40.17 MHz. This wide tuning range shows the versatility of our WGMR-based source for applications requiring adjustable spectral properties.

To further study the influence of the coupling condition on HOM interference, we measured the HOM interference in the overcoupling regime, where the photon-pair bandwidth is broadened to 43.53 MHz, as shown in Fig. \ref{fig:over}(b). Under the higher coupling rate and broader photon-pair bandwidth, the characteristic HOM beating effect remains clearly observable. The fitted beating period is 3.20 $\mu$m, which agrees well with the theoretical prediction, and is independent of the photon-pair bandwidth. This ability to flexibly control the bandwidth of photon pairs offers an additional degree of freedom for applications requiring adjustable spectral properties \cite{Zhang12,Huang25}.

%However, it is important to note that the number of modes in the generated states coupled out of the WGMR increases with the coupling rate. This multimode emission leads to a loss of indistinguishability among the interfering beams, thus reducing the achievable interference visibility. Nevertheless, the ability to flexibly control the bandwidth of photon pairs offers an additional degree of freedom for optimizing source performance. 

%\textit{Summary} 

In addition to the tunability of the photon-pair bandwidth, we note that the central wavelengths of the signal and idler photons generated from the WGMR can also be efficiently tuned. By adjusting the crystal temperature by only ten degree Celsius, the photon wavelengths can be tuned from a degenerate emission to a significantly non-degenerate condition with wavelength difference between the signal and idler photons exceeding 300 nm, which corresponds to a frequency detuning more than 90 THz. 
By further leveraging the angular phase-matching conditions in spherical geometries, photon pairs with greater nondegeneracy can be generated by selectively coupling to whispering gallery modes with higher radial mode number or cluster numbers \cite{Förtsch15}. This approach provides additional degrees of freedom for controlling the frequency detuning between the signal and idler photons, thus broadening the accessible spectral range. Moreover, the triply-resonant and highly nondegenerate PDC can be easily achieved by taking advantage of the total internal reflection within the WGMR. Our compact WGMR-based narrowband photon source does not require complex reflector fabrication or intricate cavity control.  

In summary, we demonstrate a narrowband frequency-entangled photon source based on a crystalline WGMR. 
The long coherence length of the photon pairs enables a meter-scale dynamic scanning range for HOM interferometry, while the quantum beat note provides sub-picosecond resolution. 
This enhancement makes our photon source feasible to perform time-resolved sensing for probing the coherent transfer of entangled photons \cite{Altewischer02,Cheng20} and for precise in-fiber sensing for eavesdropper localization in optical fiber networks \cite{Popp24, Chen19}. 
We also note that our bright frequency-entangled photon source may have potential applications in scalable frequency-bin-encoded quantum key distribution networks \cite{Kashi25,Tagliavacche25}. 
Implementing a ring resonator on a lithium niobate or silicon photonic integrated circuit \cite{Zhu21,Guo14} could further enable a more stable and compact realization of a frequency-entangled photon source.
Our work therefore opens new opportunities for employing narrowband frequency-entangled photon pairs in the development of novel quantum information processing technologies.
%We note that implementing a ring resonator on a lithium niobate or silicon photonic integrated circuit \cite{Zhu21,Guo14} could enable a more stable and compact realization of a frequency-entangled photon source.
%Our work thus opening up new opportunities for employing narrowband entangled photon pairs in the development of novel quantum sensing techniques.

\textit{Acknowledgments.}---
We acknowledge support from the European Commission under Project No. 101114043 — QSNP, and from the federal ministry for science, technology, and research (BMFTR) in the context of the federal government’s research framework in IT-security “Digital. Secure. Sovereign.” under Project No. 16KIS1264.

%\bibliography{flux}

\end{document}